\documentclass[structabstract]{aa}  
\usepackage{natbib} 
\bibpunct{(}{)}{;}{a}{}{,} % to follow the A&A style
\usepackage{txfonts}
\usepackage{graphicx} 

\def \HCO+{HCO$^+$} 
\def \H13CO+{H$^{13}$CO$^+$} 

\def\solar{\ifmmode_{\mathord\odot\;} \else $_{\mathord\odot}\;$\fi} % _solar
\def\mo{\ifmmode {\,{\it M}\solar\;} \else $\,M$\solar$\;$\fi}	      % M solar		
\def\am{\ifmmode{^{\scriptscriptstyle\prime}}			       % arcmin
	\else $^{\scriptscriptstyle\prime}$\fi}
\def\deg{\ifmmode{^\circ}\else$^{\circ}$\fi}			       % degree

\def\x {\ifmmode\times\else$\times$\fi}			       	       % times 

\begin{document}
   \title{Tracing gas accretion in the Galactic center using isotopic ratios\thanks{Based on observations carried out with the IRAM 30m telescope. IRAM is supported by INSU/CNRS (France), MPG (Germany) and IGN (Spain).}}

   %\subtitle{I. Overviewing the $\kappa$-mechanism}

   \author{D. Riquelme \inst{1} 
    \and M.A. Amo-Baladr\'on \inst{2} 
    \and J. Mart\'{i}n-Pintado \inst{2}
    \and R. Mauersberger \inst{3} 
    \and S. Mart\'{i}n \inst{4} 
    \and L. Bronfman \inst{5} }
    
    \offprints{D. Riquelme}
    
    \institute{Instituto de Radioastronom\'{i}a Milim\'etrica (IRAM), Av. Divina Pastora 7, Local 20, E-18012 Granada, Spain\\ 
     \email{riquelme@iram.es} 
    \and Centro de Astrobiolog\'ia (CSIC/INTA), Ctra. de Torrej\'on a Ajalvir km 4, E-28850, Torrej\'on de Ardoz, Madrid, Spain 
    \and Joint ALMA Observatory, Av. El Golf 40, Piso 18, Las Condes, Santiago de Chile, Chile
   \and European Southern Observarory, Alonso de C\'ordova 3107, Vitacura, Casilla 19001, Santiago, Chile
\and Departamento de Astronom\'{i}a, Universidad de Chile, Casilla 36-D, Santiago, Chile}

  \date{}

\abstract
% context heading (optional)
% {} leave it empty if necessary  
%{Isotopic ratio \mbox{$^{12}$C/$^{13}$C} in Galactic center is $\sim$ 20-25}
{}
% aims heading (mandatory)
{We study the \mbox{$^{12}$C/$^{13}$C} isotopic ratio in the disk of the central molecular zone and in the halo to trace gas accretion toward the Galactic center region in the Milky Way.}
% methods heading (mandatory)
{Using the IRAM 30m telescope, we observe the $J=1-0$ rotational transition of HCO$^{+}$, HCN, HNC and their $^{13}$C isotopic substitutions in order to measure the \mbox{$^{12}$C/$^{13}$C} isotopic ratio. We observe 9 positions selected throughout the Galactic center region, including clouds at high latitude; locations where the X1 and X2 orbits associated with the barred potential are expected to intersect; and typical Galactic center molecular clouds.}
% Results heading 
{We find a systematically higher \mbox{$^{12}$C/$^{13}$C} isotopic ratio ($>$40) toward the halo and the X1 orbits than for the Galactic center molecular clouds (20-25). Our results point out to molecular gas which has undergone a different degree of nuclear processing than that observed in the gas towards the inner Galactic center region.}
%conclusions
{The high isotopic ratios are consistent with the accretion of the gas from the halo and from the outskirts of the Galactic disk.}
  

   \maketitle
%________________________________________________________________
\section{Introduction}
\indent Many galaxies including our Milky Way contain huge amounts of
gas in their central few 100 pc (\citealp{Mauersberger_Henkel_1993, Morris_Serabyn_1996}). This gas
can be the fuel reservoir to feed star formation events and in many cases the central object activity (\citealp{Usero_et_al_2004,Garcia-Burillo_et_al_2005}). Both processes can occur in bursts. For example, the central region of our Galaxy, which harbors the nearest massive black hole \citep[\mbox{$\sim 4.0 \times
10^6$ M$_\odot$};][]{Ghez_et_al_2005}, may have been much more active
in the past than it is now \citep{Morris_et_al_1999}. It is obvious that such a high
activity, even if intermittent, can only be maintained if there is a
supply of gas from other regions of the Galaxy, e.g. a disk or halo,
or from intergalactic space. 

\indent Modeling of the kinematics of the inner regions of galaxies been successful in explaining inward motion,  matching the observations: \citet{Binney_et_al_1991} showed that the large-scale gas
kinematics in the Galactic center (GC) can be accounted for by a barred
galactic potential. In the dynamical model with a barred potential there
are two major families of stable prograde periodic orbits inside the
bar: the X1 orbits parallel to the bar, and the X2 orbits
orthogonal to it \citep{Contopoulos_Papayannopoulos_1980}. In the
outer parts of the bar, gas tends to follow the X1 orbits. Further
inwards, these orbits develop cusps and loops, where matter on
self-intersecting orbits gets shocked and falls toward the center,
plugging  to orbits of the X2 family \citep{Maciejewski_Sparke_2000}.
Another mechanism for gas accretion toward the
central region of galaxies is the disk-halo interaction.  A clear
example of these interactions has been recently shown by
\citet{Fukui_et_al_2006}, who found huge loop structures  (known as
``giant molecular loops"; GMLs) that connect gas at the Galactic
plane with that at higher latitudes. 

\indent However, so far there is no clear observational evidence that gas at
the outer disk (in X1 orbits) and/or at high latitudes (in the GMLs) flows toward the central regions. 
Gas at the GMLs could also be ejected to higher latitudes by winds,
originated in  past massive star-formation events  (the GC harbors giant molecular clouds with
ongoing massive star formation, e.g. in \mbox{Sgr B} and the 1.3 complex). 

Observations of gas kinematics tell us what is happening now, but do not allow to look into the past. However, studies of the isotope trace the flow since different regions in a galaxy may have their very characteristic isotopic fingerprints. This approach has the potential to reconstruct the migration history to the central part of the galaxy.\\
\indent Carbon, nitrogen and oxygen (CNO) isotopic ratios diagnostic tools for probing models of Galactic
chemical evolution \citep[see e.g.][]{Audouze_1985}. In particular,
the \mbox{$^{12}$C/$^{13}$C} isotopic ratio reflects the  history of
the gas proccesing by stars, because this ratio shows the
relative degree of primary to secondary processing in stars. While
$^{12}$C is predicted to be formed in first generation, metal-poor
massive stars, on rapid timescales, $^{13}$C is thought to be produced
primarily via CNO processing of $^{12}$C seeds from earlier stellar
generations, in a slower timescale in low and intermediate-mass stars
or novae (\citealp{Meyer_1994,
Wilson_Matteucci_1992,Prantzos_et_al_1996}). The \mbox{$^{12}$C/$^{13}$C}
isotopic ratio is one of the best-established in the GC. It presents a
clear gradient with galactocentric distance \citep[e.g.,][]{Wilson_1999},
decreasing from 80$-$90 in the solar neighborhood to 20-25  in the
inner Galaxy \citep[toward Sgr A and Sgr B, see e.g.,][]{Wannier_1980}. However, this ratio is, so far, unknown in the galactic halo and in the X1 orbits.

\indent In this paper we present observations
of the $J=1-0$ rotational transition of \HCO+, \H13CO+, HCN, H$^{13}$CN, HNC and HN$^{13}$C to derive the \mbox{$^{12}$C/$^{13}$C} isotopic ratio toward the gas in the
halo, the disk and at the kinematic components associated with the
X1 and X2 orbits in the nucleus of our Galaxy. We found very
different \mbox{$^{12}$C/$^{13}$C} isotopic ratios in the X1 orbits
and in the halo than those in the disk and in the X2 orbits. Our findings are consistent with the scenario of less processed gas in the halo and in the X1 orbits, supporting the idea of gas flowing toward the nucleus of
the Milky Way.
%____________________________________________________________________
%
\section{Observations}
\begin{figure*}
\includegraphics[width=0.9\textwidth, angle=0]{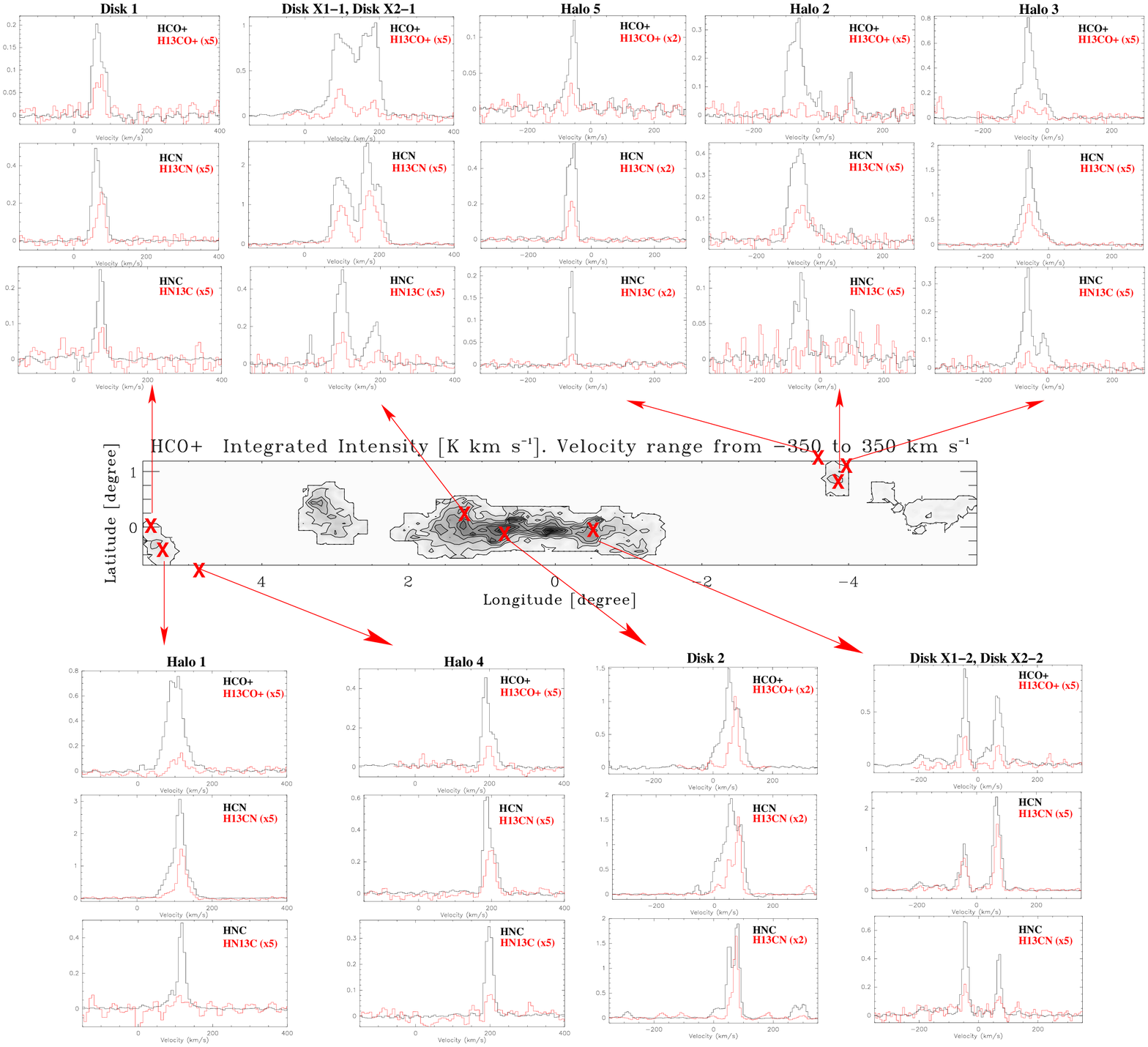}
\caption{Spectra toward selected positions in the GC in HCO$^+$, H$^{13}$CO$^+$, HCN, H$^{13}$CN, HNC and HN$^{13}$C. The positions are indicated in the HCO$^+$ integrated intensity map from \citet{Riquelme_et_al_2010}.  The $^{13}$C substitution is scaled by a factor of 2 or 5 to allow an easy visualitation of the spectra. The factor is indicated in each spectra. }
\label{Fig1}
\end{figure*}

\indent Observations were carried out with the IRAM-30m telescope at
Pico Veleta (Spain) in June and December 2009. We used the E090 band of the new EMIR receiver, which provide a bandwidth of $\sim 8$ GHz  (from 83.7 to 91.1 GHz)
simultaneously in both polarizations. For the backend, we used the WILMA
autocorrelator, providing a resolution of 2 MHz or 6.8
km/s. Observations were performed in position switching mode where the reference off-positions were checked to be free of emission (Table \ref{pos_off} list the emission-free positions).  The
pointing was checked every 2 hours against the source 1757-240
providing an accuracy better than $5''$. Data were calibrated using
the standard dual load system. In this work, we use the antenna
temperature scale $T^*_{\rm A}$. Main beam temperatures, $T_{\rm MB}$, can
be obtained using $T_{\rm MB}=Feff/Beff\times T^*_{\rm A}$, where the forward efficiency is $Feff=95\%$ and the main
beam efficiency is $Beff = 81\%$ at $86$ GHz. All positions
were observed for no less than 30 min  providing an rms noise antenna
temperature of about 6 mK in the weakest lines (e.g. \H13CO+).\\
\indent We observed the nine positions shown in Table \ref{position} displayed in the
large scale map of the HCO$^+$ emission shown in Fig. \ref{Fig1}. Five of the observed
positions are located in the GMLs (Halo in Table \ref{position}, Table \ref{res} and
Fig. \ref{Fig1}): Three of them are located at the foot points and two
at the top of the loops. Two positions are in the disk toward the locations
of the expected interactions between the X1 and X2 orbits
(Disk$\,$X1 and Disk$\,$X2  in Table \ref{position}, Table \ref{res} and Fig. \ref{Fig1}). A
pair of positions toward the Galactic plane (Disk in Table \ref{position}, Table \ref{res} and Fig. \ref{Fig1}) have been used as
reference measurements.  The positions at the foot points of the GMLs
were selected from the intensity peaks of the SiO maps of \citet{Riquelme_et_al_2010} and previous results from higher angular resolution mapping
with the Mopra-22m telescope (Riquelme et al., in prep.). The \citet{Fukui_et_al_2006} maps of the GMLs were used to select the
position at the top of the loop. The positions in the 1$\deg$.3 and
in Sgr C complexes correspond to the locations where interactions
between the X1 and X2 orbits are expected (see,
\citealp{Binney_et_al_1991, Stark_et_al_2004}). CO and
SiO(2-1) maps of these regions (\citealp{Tanaka_et_al_2007}, Amo-Baladr\'on et al., in prep.)  were used for final selection of the positions. The control points in the disk are located in Sgr B2 and in the  l=5\deg.7 complex selected from \citet{Martin_et_al_2008} and \citet{Fukui_et_al_2006}, respectively.
\begin{table*}
\caption{Galactic and equatorial coordinates of the emission-free reference positions. \label{pos_off}}
{\small
\begin{center}
\begin{tabular}{cccc}
\hline\hline
%\\
\multicolumn{2}{c}{Galactic coordinates}& \multicolumn{2}{c}{Equatorial coordinates} \\
$l$ [\deg]& $b$ [\deg] &$\alpha_{J2000}$ & $\delta_{J2000}$ \\ \hline
5.75  &  1.0     & 17:54:56.6 & $-$23:29:13 \\ %refClumpD
4.5   & $-1.29$  & 18:00:57.3 & $-$25:43:12 \\ %refToploop3
356.375&   1.5   & 17:30:48.0 & $-$31:11:48\\%refClumpC
359.75 & $-0.25$ & 17:46:00.1 & $-$29:16:47 \\ %refSgrC
1.0   & $-1.0$   & 17:51:52.9 & $-$28:35:41\\%ref1.3complex
0.65  &  0.2     & 17:46:23.0 & $-$28:16:37     \\%refSgrB
\end{tabular}
\end{center}}
\end{table*}
\begin{table*}
\caption{Observed positions. \label{position}}
{\small
\begin{center}
\begin{tabular}{lccccl}
\hline\hline
%\\ 
Associated  & \multicolumn{2}{c}{Galactic coordinates}& \multicolumn{2}{c}{Equatorial coordinates} & this work \\
object      &  $l$ [\deg]      & $b$ [\deg]           & $\alpha_{J2000}$ & $\delta_{J2000}$      &   \\ \hline    
M$+5.3-0.3$ & $5.45$           &  $-0.324$            & $17^{\rm h}59^{\rm m}17.8^{\rm s}$ &$-24\deg24'38''$&Halo\,1  \\
M$-3.8+0.9$ & $356.206$        &  $0.83$              & $17^{\rm  h}32^{\rm m}59.8^{\rm s}$ &$-31\deg42'17''$&Halo\,2  \\
M$-3.8+0.9$ & $356.179$        & $0.925$              & $17^{\rm h}32^{\rm m}33.2^{\rm s}$ &$-31\deg40'31''$&Halo\,3  \\
Top Loop    & $4.75$           &  $-0.8$              & $17^{\rm h}59^{\rm m}34.9^{\rm s}$ &$-25\deg15'16''$  &Halo\,4\\
Top Loop    & $356.549$        & $1.339$              & $17^{\rm h}31^{\rm m}52.5^{\rm s}$ &$-31\deg08'19''$ &Halo\,5\\
1.3 complex & $1.28$           &  $ +0.07$            & $17^{\rm h}48^{\rm m}21.9^{\rm s}$ &$-27\deg48'19''$&Disk\,X1-1, Disk\,X2-1\\
Sgr C       &$359.446$         &  $-0.124$            & $17^{\rm h}44^{\rm m}46.9^{\rm s}$ &$-29\deg28'25''$&Disk\,X1-2, Disk\,X2-2\\
Galactic plane at $l\sim 5\deg.7$  &$5.75$& $0.25$        & $17^{\rm h}57^{\rm m}46.5^{\rm s}$ &$-23\deg51'51''$&Disk\,1   \\
Sgr B2      & $0.6932$         &  $ -0.026$           & $17^{\rm h}47^{\rm m}21.9^{\rm s}$ &$-28\deg21'27''$&Disk\,2   \\
\end{tabular}
\end{center}}
\end{table*}
%%__________________________________________________________________
\section{Results}
\indent Fig. \ref{Fig1} shows the spectra taken in all the observed positions in the J=1$\to$0 transition of the main isotope of \HCO+, HCN and HNC and their $^{13}$C isotopologues. All the species have been clearly detected in all our sources, except HN$^{13}$C, with tentative detections in ``Halo\,1'' and ``Halo\,3'', and a non detection in ``Halo\,2''.

\indent Table \ref{res} shows the integrated line intensity ratios between the different isotopologues derived for all positions and velocity components. To distinguish between the different kinematical components, the line intensity ratios were obtained by integrating the line profiles in the velocity ranges given in Table \ref{res}. We derived line intensity ratios from three different species, obtained the largest values from \HCO+.

\begin{table*}
\caption{Intensity ratios from the observed $^{12}$C and $^{13}$C
isotopomers. \label{res}} 
{\small
\begin{center}
\begin{tabular}{c|c|c|c|c|c}
\hline\hline
%\\ 
Source    & Velocity Component  & Velocity Range &  HCO$^+$/H$^{13}$CO$^+$ & HCN/H$^{13}$CN & HNC/HN$^{13}$C  \\
          &   LSR [km\,s$^{-1}$]&    [km\,s$^{-1}$]      &ratio of $\int\rm{T_A^*dv}$&ratio of $\int\rm{T_A^*dv}$& ratio of $\int\rm{T_A^*dv}$\\\hline\hline
%\\ 
Halo\,1   &        100       & [50, 190]      & 45.5$\pm$ 5.4 & 13.5$\pm$0.2 & $\geq$ 25.6\\
          &        87        & [50, 97]       &$\geq$73.9     & 25.8$\pm$2.0 & $\geq$7.5\\
          &        117       & [97,135]       & 32  $\pm$ 3.7 & 11.3$\pm$0.1 & $\geq$37.5 \\
          &        144       & [135,190]      & 39.1$\pm$24.7 & 16.7$\pm$1.8 & $\geq$2.7\\
%\\				     			   	    	     	     
Halo\,2   &       $-62$      & [$-115$,$-20$]  &73.1$\pm$36.5  &14.6 $\pm$1.0 & $\geq$15.4\\
          &   left  wing     & [$-115$,$-70$]  &$\geq$34.4     &21.2 $\pm$3.3 & $\geq$8   \\
          &   right wing     & [$-70$, $-20$]  &53.2$\pm$26.1  &11.6 $\pm$0.9 & $\geq$13.6\\
%\\				     			   	    	     	     
Halo\,3     &     $-60$      & [$-120$,$-30$]  &38   $\pm$5.0  &13   $\pm$0.3 & $\geq$40.1\\
          &   left  wing     & [$-120$,$-80$]  &54.2 $\pm$37.7 &19.2 $\pm$1.9 & $\geq$10.4\\
          &   central peak   & [$-80$, $-30$]  &35.7 $\pm$4.0  &12   $\pm$0.2 &48.3 $\pm$ 21.8\\
%\\
%\\
Halo\,4   &       200        & [150, 250]     & 28.3 $\pm$5.4   &11.8 $\pm$0.9  & 14.9 $\pm$3.1 \\
          &       peak       & [150, 210]     & 29.2 $\pm$7.5   &11.7 $\pm$0.8  & 17.9 $\pm$3.9 \\
          &   right wing     & [210, 250]     &$\geq$10.6       &12.4 $\pm$4.0  &  7.1 $\pm$3.4 \\
%\\%\\				     			   	    	     	     
Halo\,5   &      $-50$       & [$-100$,$-40$]  &13.8 $\pm$5.0 & 6.9 $\pm$0.3 & 22.8 $\pm$12 \\
%\\				     			   	    	     	     
Disk\,X1-1&       180       & [140,230]        &56   $\pm$ 6.4 & 10.8 $\pm$0.1  &25.5 $\pm$ 7.5\\
           &  left wing      & [140,180]       &57.2 $\pm$ 7.5 & 11.6 $\pm$0.1  & $\geq$8\\
           &  right wing     & [180,230]       &54.4 $\pm$ 11  & 9.9  $\pm$0.2  &15.7 $\pm$ 4.4\\
%\\				     			   	    	     	     
Disk\,X2-1&        95       & [50,140]         &29   $\pm$ 1.6 & 12.1 $\pm$ 0.2 &22.1 $\pm$ 2.7\\
           &  left wing      & [50,92]         &32.4 $\pm$ 3.7 & 13.9 $\pm$ 0.4 &22.8 $\pm$ 4.0\\
           &  right wing     & [92,140]        &27.3 $\pm$ 1.7 & 11.3 $\pm$ 0.2 &21.7 $\pm$ 8.6\\
%\\   						      	   	                			      
Disk\,X1-2&       67        & [0,100]         & 42.1$\pm$ 8.6 & 9.4 $\pm$ 0.1 &9.9 $\pm$ 1.3\\
%\\   						      	   	                			      
Disk\,X2-2&       $-43$     & [$-80$,$-20$]   & 21.7 $\pm$1.9 & 6.6 $\pm$0.1 & 13.1 $\pm$1.0 \\
%\\   						      	   	                			      
Disk\,1   &       66         & [35,105]       &14.0 $\pm$2.8  & 13.1 $\pm$ 1.0 & 15.8 $\pm$ 3.6\\
          &   central peak   & [35,70]        &16.4 $\pm$4.6  & 23.1 $\pm$ 3.8 & $\geq$11.4\\ 
          &   right wing     & [70,105]       &11.6 $\pm$3.4  &  7.8 $\pm$ 0.7 & 11.7 $\pm$ 2.7\\
%//   
Disk\,2   &   left wing      & [0, 43]       &$\geq$29.1      & 9.1  $\pm$ 0.2 &$\geq$21.9\\
          &       55         & [43, 97]      &4.1  $\pm$ 0.1 & 3.5  $\pm$ 0.1 & 3.7 $\pm$ 0.1\\
          &   right wing     & [97,135]      & 16.1 $\pm$ 2.3 & 13.8 $\pm$ 1.3&$\geq$2.6\\ \hline
\end{tabular} 
\end{center}}
\end{table*}
%%
%__________________________________________________________________
\section {The $^{12}$C/$^{13}$C isotopic ratios \label{12C/13C}}
\indent The isotopologues that we observed in this work have very similar rotational constant and Einstein coefficients. Therefore the beam size is very similar and for optically thin emission, one would expect that line intensity ratios would be directly converted into column density ratis, i.e., 
%\begin{eqnarray}
%\int {\rm T_{H^{12}CO^+}\rm dv}/\int {\rm T_{H^{13}CO^+} dv}&=&{\rm N( H^{12}CO^+})/{\rm N( H^{13}CO^+})\nonumber \\
%             &=&\rm [H^{12}CO^+]/[H^{13}CO^+]\nonumber ,
%\end{eqnarray}
$\int {\rm T_{H^{12}CO^+}\rm dv}/\int {\rm T_{H^{13}CO^+} dv}={\rm N( H^{12}CO^+})/{\rm N( H^{13}CO^+})=\rm [H^{12}CO^+]/[H^{13}CO^+]$ 
%\indent The isotopologues that we observed in this work have very
%similar rotational constants and Einstein coefficients. Therefore the
%beam size is very similar and for optically thin emission, one would
%expect that line intensity ratios would be directly converted into
%column density ratios, i.e., $\int {\rm T_{H^{12}CO^+} \rm dv}/\int
%{\rm T_{H^{13}CO^+} dv}={\rm N( H^{12}CO^+})/{\rm N( H^{13}CO^+})=\rm
%[H^{12}CO^+]/[H^{13}CO^+]$

\indent In general, the molecular isotopologue ratios (see Table
\ref{res}) do not translate  directly into \mbox{$^{12}$C/$^{13}$C}
isotopic ratios. Opacity, chemical fractionation and selective
photodissociation effects must be considered to derive the isotopic
ratios from molecular line intensity ratios. In the following, we will
discuss the importance of these effects on the
\mbox{$^{12}$C/$^{13}$C} isotopic ratios that we derive from our
integrated intensity ratio of the isotopologues.
\subsection{Opacity effects, isotopic fractionation and selective photodissociation} 
\indent In our case, it is not possible to estimate the line optical depth from the
emission of just one single transition. Since optical depth effects will saturate the emission from the most abundant isotopologue, our derived molecular isotopic ratios must be considered to be lower limits to the actual \mbox{$^{12}$C/$^{13}$C} isotopic ratios.
 In almost all cases, the HCO$^+$/H$^{13}$CO$^+$ intensity ratios or their limits are higher than those derived from the HCN and HNC isotopomers. This may be interpreted as HCO$^+$ being the least optically thick among the three molecular line emission used in this work. Therefore, HCO$^+$/H$^{13}$CO$^+$ intensity ratios give the most stringent limits to the \mbox{$^{12}$C/$^{13}$C} isotopic ratios. We still cannot exclude that also \HCO+ are affected by opacity effects. Hence \mbox{$^{12}$C/$^{13}$C} isotopic ratios are even higher than inferred from the HCO$^+$/H$^{13}$CO$^+$ intensity ratios. In the following we will therefore only consider the isotopologue ratios derived from \HCO+.

\indent Another important effect may be the chemical fractionation
\citep[][and references therein]{Wilson_1999}. \citet{Langer_et_al_1984} have studied the
fractionation of carbon and oxygen isotopes with a  time-dependent
chemical model. Their model considers a cloud lifetime of $10^8$ years,
temperatures from 6 to 80 K, H$_2$ densities from $5\times 10^2$ to
$1\times 10^5$  cm$^{-3}$, and a wide range of metal abundances.
Fractionation in \HCO+ was also studied by \citet{Woods_Willacy_2009}
for protoplanetary disks, which can have a larger impact in the
\mbox{$^{12}$C/$^{13}$C} ratio, but with temperatures and densities
different from that of the GC. \citet{Langer_et_al_1984} found that the behavior of the carbon isotope ratios can be split into three groups: CO, \HCO+, and ``carbon-isotope pool'' (which includes the remaining carbon species, such as C$^+$, H$_2$CO, CS, etc). They found that the $^{13}$C is enhanced in the CO (specially at low temperature, low density, and high metal abundance), the $^{12}$C is enhanced in the ``carbon isotope pool'' group, and the \HCO+ could present both fractionation effects, depending on the physical condition. The behavior of \HCO+ is related to the formation of this molecule from both CO and from the ``carbon pool isotopes''. The formation of \HCO+ has been explained by the ion-molecule
chemistry \citep{Wilson_2009},  where the reaction:
\begin{equation}
\label{prodHCO}
 \rm H_3^+ + \rm CO  \leftrightarrow  \rm HCO^+ + \rm H_2
\end{equation}
likely leads the production of \HCO+. \H13CO+
is more tightly bound than \HCO+ by 0.8 meV (9 K). The reactions that
have the potential to produce isotopic fractionation occurs by
proton-switching reactions between formyl ions and carbon monoxide by
(\citealp{Langer_et_al_1984,Langer_et_al_1978})
\begin{equation}
\label{fractHCO}
 \rm HCO^+ + ^{13}\rm CO  \leftrightarrow  \rm H^{13}CO^+ + ^{12}\rm CO+ 9 \rm K
\end{equation}
\indent As it is shown in Fig. 1 of \citet{Langer_et_al_1984}, the isotopic ratio derived from \HCO+ presents only moderate fractionation. The \mbox{H$^{12}$CO$^+$/H$^{13}$CO$^+$} ratio is slightly enhanced at low density and moderate temperature, and decreases at low temperatures and certain densities (n(H$_2)<10^3$ and n(H$_2)>10^{4.5}$). In this model, \citet{Langer_et_al_1984} considered the formation paths of \HCO+ shown in the Table 2 of \citet{Graedel_et_al_1982}. \citet{Huettemeister_et_al_1998}
and \citet{Rodriguez-Fernandez_et_al_2002} have found that the gas
in the GC has temperatures ranging from 20 to 200 K.  For this
temperature range, the predicted fractionation of \HCO+ is negligible and the fractionation of $^{13}$CO cannot occur. In the worst case, the isotopic ratio throughout the GC should not increase selectively one of the molecular isotopologue against the other by more than
20$\%$. 

\indent For molecular clouds affected by UV radiation, selective
photodissociation can take place, which would increase the
\mbox{$^{12}$C/$^{13}$C} isotopic ratio. The more abundant molecules (the main isotopologues) are less affected by photodissociation through self-shielding against UV radiation than the rarer isotopologues.
 So far, studies of the \HCO+ and its
$^{13}$C isotopologue in PDRs have not shown any clear evidence for
selective photodissociation \citep{Fuente_et_al_2003}. \citet{Milam_et_al_2005} concluded that the
\mbox{$^{12}$C/$^{13}$C} isotopic ratio derived from high density
tracers like CN are very unlikely influenced by isotope-selective
photodissociation. However the self-shielding and photodissociation of CO \citep[e.g.,][]{Glassgold_et_al_1985,Bally_Langer_1982, Chu_Watson_1983} could affect our derived \mbox{$^{12}$C/$^{13}$C} ratios through the formation of \HCO+ via the reaction in Eq. \ref{prodHCO}. Modeling this effect, \citet{Chu_Watson_1983} concluded that the selective photodissociation will have a negligible impact in the \mbox{$^{13}$CO/$^{12}$CO} ratio. Although the local enviroment of the GMLs is so far unknown, there are no
signposts of UV radiation. There is only a nearby ultracompact H\ II region identified at $(l,b)\sim(356\deg.25, 0\deg.7)$,
but its radial velocity of $\sim 120$ km\,s$^{-1}$, suggests that it is not
associated with the foot points \citep{Torii_et_al_2010}, making very
unlikely any effect in our derived isotopic ratios due to
photodissociation. The small \mbox{$^{13}$CO/$^{12}$CO} ratio observed in the GMLs by \citet{Torii_et_al_2010} do not support that the bulk of the gas is self shielded in $^{12}$CO but photodissociated in $^{13}$CO. The small ratios are likely due to optically thick emission in $^{12}$CO, suggesting rather large CO column densities. Then the bulk of the gas will not be affected by the rather low fluxes of Far UV radiation inferred from the lack of H\ II regions \citep{Torii_et_al_2010} in the area. Under these conditions, it is highly unlikely that the fractionation/self-shielding effect in the \mbox{$^{13}$CO/$^{12}$CO} could explain the observed high \mbox{H$^{12}$CO$^+$/H$^{13}$CO$^+$} ratios. We conclude that the line intensity ratios
derived from our data correspond to a lower limit to the actual
\mbox{$^{12}$C/$^{13}$C} isotopic ratio.

\indent We derive \mbox{H$^{12}$CO$^+$/H$^{13}$CO$^+$} intensity ratios from 4 to $>74$.
The lowest value of 4 is found toward the source Disk\,2, in the
position of Sgr B2 where one would expect a value of $\sim 25$. Such a low measured value
indicates that \HCO+ is optically thick. From the results in Table \ref{res} we find a systematic trend in the isotopologue ratios with the largest values, $>74$,  found toward Halo\,1 and Halo\,2. Also Halo\,3 and Disk\,X1-1 and Disk\,X1-2 show large values of $>42$. 

%__________________________________________________________________
\section{Discussion}
\indent It is remarkable that there is a systematic trend of the 
\mbox{$^{12}$C/$^{13}$C} isotopic ratio found toward the halo and in the disk for the kinematic components
associated with the X1 orbits, to be systematically larger by a factor of at least 2 than the standard
value in the GC of $\sim$ 25. This suggests that there is less processed material in
the halo and in the X1 orbits than in the X2 orbits and the reference GC positions in the disk.

\indent Surprisingly, the isotopic ratios measured in the Halo\,4 and Halo\,5 positions at the top of the GMLs show a lower $^{12}$C/$^{13}$C isotopic ratio than in the other halo positions, closer to the ``standard'' values of the disk. From our data, it is still not clear if this low isotopic ratio is due to opacity effects in the main isotope or due to a mixture with more processed gas from stellar nucleosynthesis in the GC region, likely ejected to high latitudes from the disk.

\subsection{History of the chemical evolution of the molecular gas in the Galactic center}
\indent The lower limits to the \mbox{$^{12}$C/$^{13}$C} isotopic abundance ratios derived from our \HCO+ data toward the 
``typical'' GC molecular clouds (Disk\,1, Disk\,2, Disk\,X2-1, Disk\,X2-2) range between 4 to 32, with
an average value of 20. These results agree with the values found in
the literature for the GC \citep[20-25,][]{Wilson_1999}, and with the average value derived by \citet{Riquelme_et_al_2010} of $19.8$ (from the \HCO+/\H13CO+ integrated intensity ratio throughout the GC region), showing a considerable nuclear processing. In contrast, we have found much higher limits to the isotopic ratios toward the locations where gas in the inner disk\footnote{Throughout this work, all the positions observed and discussed are in the Galactic center region (in the central kpc of the Galaxy). We call ``disk'' to the disk in the CMZ, ``inner disk'' to the allowed velocities in the X2 orbits, and ``outer disk'' and ``outskirts of the disk'' to the non-circular motions (X1 orbits).}  is expected to be interacting with that in the halo and in the disk. 
Indeed, the gas in the X1 orbits is of different nature than that in the X2 orbits, as clearly reflected by their different isotopic ratios of $\sim 55$ and $\sim 21-30$, respectively. Furthermore, in the halo, the interaction occurs
when the gas in the top of the loops flows toward the foot points of the GMLs. In
both cases, the isotopic ratio reflects a different nature of the gas
than that in the GC. The \mbox{$^{12}$C/$^{13}$C} isotopic ratio of the gas in the halo and in the X1 orbits are in general larger than 40, reaching values larger than 70 in ``Halo\,1''.
\indent Previous studies show a clear gradient in the
\mbox{$^{12}$C/$^{13}$C} ratio with galactocentric distance. The ratio
changes from a value of about 50 in the inner Galaxy (4 kpc) to nearly
70 in the local ISM, and $\sim 90$ in the solar system,  with a value
$\sim 20-25$ in the GC \citep{Wilson_1999}, which reflects higher
nuclear processing toward the inner Galaxy. Studies
of the isotopic ratios in the GC (\citealp{Henkel_et_al_1985,
Stark_1981,Langer_Penzias_1990, Wannier_1980}) have shown that the
interstellar gas at the GC is in an advanced state of chemical
evolution (which corresponds to an enrichment of $\sim 3-4$ with
respect to the local gas).  
This gradient indicates that the material is flowing to the 
inner central kiloparsec of the GC from the Galactic halo (in the case of
GMLs) and the outskirts of the Galactic disk (in case of X1-X2
interacting orbits).
\subsection{The origin of molecular gas in the GML}
\indent Our lower limits to the \mbox{$^{12}$C/$^{13}$C} isotopic ratios in the halo sources for the
foot points of the GMLs  are close to the values measured in the local
ISM \citep[69$\pm 6$,][]{Wilson_1999}. Based on energetical and morphological arguments, \citet{Fukui_et_al_2006} argued
that it is impossible that the loop features could be created by
supershells or supernova explosions. Our data strongly support their claim.
If the loop features were formed by supernova or hypernova explosions,
the gas at the foot points should have been ejected from the GC to
high latitudes, reflecting the isotopic ratio found throughout the
GC. It is likely that the gas in the GMLs have been accreted
from high latitudes. However, if the low isotopologue ratio found at the top of the loops is confirmed to be the actual isotopic ratios, this might be an indication that some gas in the halo could have been also ejected from the disk. \\
\indent Magneto-hydrodynamical
simulations have been successful in explaining the formation of the
loops (\citealp{Matsumoto_et_al_1988,
Horiuchi_et_al_1988,Fukui_et_al_2006, Machida_et_al_2009,
Takahashi_et_al_2009}). They claimed that the GMLs have been formed as
a natural consequence of a differentially rotating magnetized gas disk
under a strong gravitational potential \citep{Machida_et_al_2009}.  It is
found that the loop efficiently accumulates gas generating a dense
layer at the top of the loops. However, our results are not consistent
with the magneto-hydrodynamical loop scenario in which the gas in the GML 
falling to their foot point should share the same isotopic ratios
as the gas in the Galactic disk. \citet{Morris_2006} has also pointed out
that there is too much gas left at the top of the loops if the gas were
flowing down along the magnetic field lines. \citet{Morris_2006} and
\citet{Torii_et_al_2009a} alternatively proposed that ambient HI gas may be
converted in the GML into H$_2$ during flotation. It is likely that the loops are surrounded by H\,I gas. The rising portion of the magnetic loops suffer shocks that compress the relatively rarefied atomic gas in front of
it, leading to rapid cooling and, ultimately, to a phase transition
from atomic to molecular gas \citet{Morris_2006}. Therefore the molecular gas that defines
the loops is constantly replenished. This scenario also support our results, because the gas in the foot point of the loop has not yet been processed by star formation.
\subsection{Gas accretion from the outer disk}
\indent We have also found high isotopic
ratios toward the non-circular components in the nuclear disk which
have been explained in terms of a stellar bar driven potential
\citep{Binney_et_al_1991}. The 1\deg.3 complex has been observed and
studied by several authors (\citealp{Tanaka_et_al_2007,
Oka_et_al_2001, Huettemeister_et_al_1998}). This molecular complex
shows broad velocity widths and a large latitudinal height scale. The
molecular  emission shows two velocity components, one centered at
$\sim 110$ km\,s$^{-1}$ and the other at $\sim 190$ km\,s$^{-1}$, which correspond to our
Disk\,X2-1 and Disk\,X1-1 sources
respectively. \citet{Tanaka_et_al_2007} found an enhancement of the CO
$J=3-2$, \HCO+ $J=1-0$, HCN $J=1-0$ with respect to the CO $J=1-0$, and an
enhancement in the abundance of  SiO in the high
velocity component ($>110$ km\,s$^{-1}$). They also identified several
expanding shells very  prominent at $v_{\rm LSR} < 110$ km\,s$^{-1}$ associated with
recent star formation. They conclude that the two components
correspond to different types of gas with different kinematics and
physical conditions. Our Disk\,X1-1 component presents an
enhancement of the SiO/$^{13}$CO intensity ratio of a factor $\sim 3$ \citep{Tanaka_et_al_2007}
with respect to the lower velocity component (Disk\,X2-1). Like
\citet{Huettemeister_et_al_1998}, \citet{Tanaka_et_al_2007} explained
the high SiO abundance as a consequence of recent shock activity.

\indent From our measurements of isotopic ratios, we conclude  that the gas
associated with the lowest velocity component presents a ``typical'' GC 
isotopic ratio consistent with  previous ideas of chemistry dominated
by shocks generated by supernova explosions. It is very unlikely, however, that
the higher velocity component, Disk\,X1-1, could be associated to
the supernova scenario like the Disk\,X2-1 component, as also
inferred from the high isotopic ratio found toward this
source. The most likely scenario is that the high SiO abundance is the
result of the shocks generated by the transfer of gas between the X1
and X2 orbits as suggested by the potential bar scenario.
%______________________________________________________________
\section{Conclusions}
  We have determined lower limits to the $^{12}$C/$^{13}$C isotopic ratio toward 9 positions in the GC region, both in the Disk and in the Halo. In contrast with the values found in the disk sources ($\sim 4-30$), we found high isotopic ratios toward the locations where the disk-halo ($\sim >40- >70$) and X1-X2 orbit interactions ($>42->56$). Our results are consistent with a scenario where gas from the halo is accreted to the disk and with the transfer of gas from the outskirt of the disk to the GC through X1 and X2 orbits as suggested by the potential bar scenario.
\begin{acknowledgements}
D.R. and R.M. were supported by DGI grant AYA 2008-06181-C02-02. J.M-P. and S.M. have been partially supported by the Spanish MICINN under grant number ESP2007-65812-C02-01. L.B. acknowledges support from FONDAP Center for Astrophysics 15010003 and from Center of Excellence in Astrophysics and Associated Technologies (PFB 06).
\end{acknowledgements}
\bibliographystyle{aa} % style aa.bst
\bibliography{referencias} % your references Yourfile.bib

\clearpage

\begin{figure*}
\includegraphics[width=0.8\textwidth, angle=0]{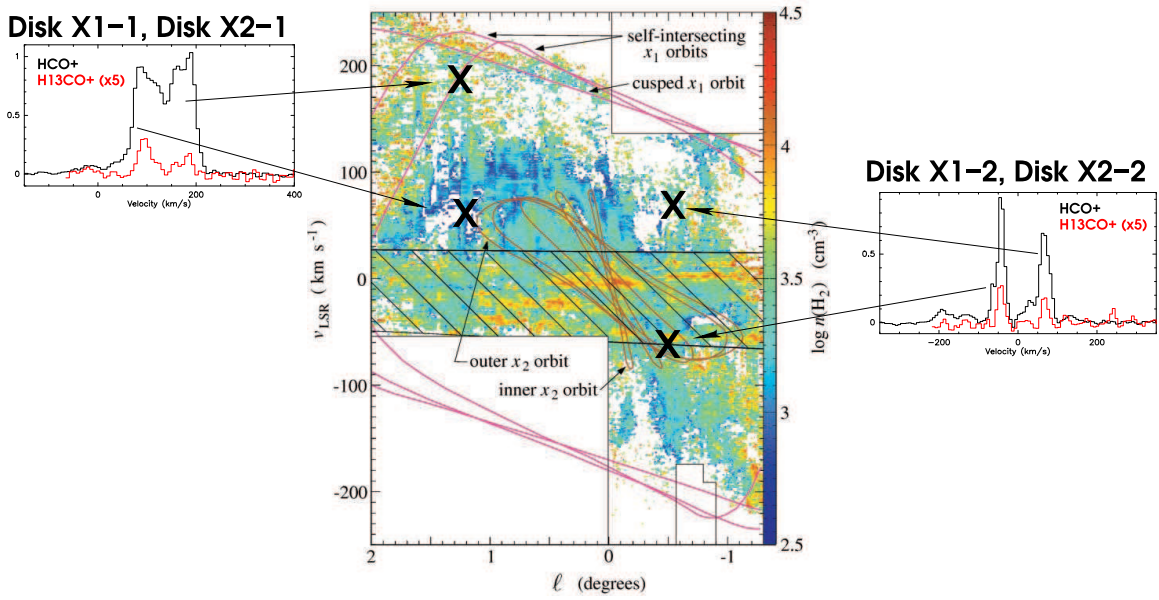}
\caption{Spectra toward the X1 and X2 orbits in HCO$^+$ and H$^{13}$CO$^+$ (scaled by a factor of 5). Left: spectra toward the 1.3 complex. Center:l-v diagram showing the X1 and X2 orbits superimposed \citep{Stark_et_al_2004}. Right: spectra toward Sgr C region.}
\label{Fig2}
\end{figure*}
%\clearpage
%______________________________
\end{document}